\documentclass{report}

\usepackage[utf8]{inputenc}
\usepackage{authblk}
\usepackage{endnotes}
\usepackage{import}
\usepackage{mathtools}
\usepackage{url}
\usepackage[hidelinks]{hyperref}
\usepackage{natbib}
\usepackage{csquotes}

\hypersetup{pdfauthor={Paul Waddell;Geoff Boeing;Max Gardner;Emily Porter},
			pdftitle={An Integrated Pipeline Architecture for Modeling Urban Land Use, Travel Demand, and Traffic Assignment},
			pdfsubject={Land Use and Transportation Modeling},
			pdfkeywords={land use;travel behavior;traffic assignment;activity-based modeling;simulation;networks;urban planning;regional planning;UrbanSim;ActivitySim;high-performance computing}}

\title{An Integrated Pipeline Architecture for Modeling Urban Land Use, Travel Demand, and Traffic Assignment\footnote{Technical report for U.S. Department of Energy SMART Mobility Urban Science Pillar: Coupling Land Use Models and Network Flow Models}}

\author[1]{Paul Waddell}
\author[1]{Geoff Boeing}
\author[2]{Max Gardner}
\author[2]{Emily Porter}
\affil[1]{Department of City and Regional Planning, University of California, Berkeley}
\affil[2]{Department of Civil and Environmental Engineering, University of California, Berkeley}

\date{January 2018}

\begin{document}

\maketitle
\let\footnote=\endnote
\renewcommand{\thesection}{\arabic{section}}

\begin{abstract}
Integrating land use, travel demand, and traffic models represents a gold standard for regional planning, but is rarely achieved in a meaningful way, especially at the scale of disaggregate data. In this report, we present a new pipeline architecture for integrated modeling of urban land use, travel demand, and traffic assignment. Our land use model, UrbanSim, is an open-source microsimulation platform used by metropolitan planning organizations worldwide for modeling the growth and development of cities over long ($\sim$30 year) time horizons. UrbanSim is particularly powerful as a scenario analysis tool, enabling planners to compare and contrast the impacts of different policy decisions on long term land use forecasts in a statistically rigorous way. Our travel demand model, ActivitySim, is an agent-based modeling platform that produces synthetic origin--destination travel demand data. Finally, we use a static user equilibrium traffic assignment model based on the Frank-Wolfe algorithm to assign vehicles to specific network paths to make trips between origins and destinations. This traffic assignment model runs in a high-performance computing environment. The resulting congested travel time data can then be fed back into UrbanSim and ActivitySim for the next model run. This technical report constitutes the FY18 Q1 deliverable for the U.S. Department of Energy SMART Mobility Urban Science Pillar task 2.2.2.2018: Coupling Land Use Models and Network Flow Models. It introduces this research area, describes this project's achievements so far in developing this integrated pipeline, and presents an upcoming research agenda.
\end{abstract}

\tableofcontents
\newpage

\section{Introduction}
\label{sec:intro}
\subsection{Need for an integrated modeling pipeline}

The overarching objective of this project is to develop an integrated modeling pipeline that encompasses land use, travel demand and traffic assignment to model the combined and cumulative impacts of transportation infrastructure and land use regulations. A key motivation for developing such a model system is that the urban environment
is complex enough that it is not feasible to anticipate the effects of alternative infrastructure investments and land use policies without some form of analysis that could reflect the cause and effect interactions that could have both intended and unintended consequences.

Consider a highway expansion project, for example. Traditional civil engineering training from the mid 20\textsuperscript{th} century suggested that the problem was a relatively simple one: excess demand meant that vehicles were forced to slow down, leading to congestion bottlenecks. The remedy was seen as easing the bottleneck by adding capacity, thus restoring the balance of capacity to demand. Unfortunately, as Downs (2004) has articulately explained, and most of us have directly observed, once capacity is added it is quickly used, leading some to conclude that \enquote{you can't build your way out of congestion}.

Highway congestion is a difficult problem because individuals and organizations adapt to changing circumstances. When new capacity is available vehicle speeds initially increase, but this drop in travel time on the highway allows drivers taking other routes to change to this now-faster route, or to change their commute to work from a less-convenient shoulder of the peak time to a mid-peak time, or to switch from transit or car-pooling to driving alone, adding demand at the most desired time of the day. Over the long-term, developers take advantage of the added capacity to build new housing and commercial and office space, households and firms take advantage of the accessibility to move farther out where sites are less expensive.

The highway expansion example illustrates a broader theme: urban systems, including the transportation network, the housing market, the labor market (commuting), and other real estate markets are closely interconnected, much like the global financial system. An action taken in one sector ripples through the entire system to varying degrees, depending on how large the intervention is and what other interventions are occurring at the same time.

This brings us to a second broad theme: interventions are rarely coordinated with each other, and often are conflicting or have a compounding effect that was not intended. This pattern is especially true in metropolitan areas consisting of many local cities and possibly multiple counties - each of which retain control of land use policies over a fraction of the metropolitan area, and none of which have a strong incentive, nor generally the means, to coordinate their actions. It is often the case that local jurisdictions are taking actions in strategic ways that will enhance their competitive position for attracting tax base-enhancing development and residents. It is also systematically the case that transportation investments are evaluated independently of land use plans and the reactions of the real estate market.

In order to better support the analysis of the impacts of transportation infrastructure and land use regulations within large and complex urban regions, we propose to develop an integrated pipeline for modeling urban land use, travel demand and traffic assignment and to compute transportation-related energy consumption. The project lends itself to further extension to address building energy consumption as well, creating the potential to coherently simulate transport and building energy demand for the first time in a coherent way, at an urban and metropolitan scale.

\subsection{Overview of pipeline architecture}

Three models are integrated within this project. UrbanSim is a model system developed to represent long-term dynamics of urban development and its interaction with transportation systems. ActivitySim is an activity-based travel demand model system, and it was developed using the UrbanSim platform as its starting point. The third model component is a static user equilibrium traffic assignment model using a standard Frank-Wolfe algorithm.  All three model systems are open source and implemented using the Python programming language, enabling broad collaboration within the research community and the public agencies who could benefit from its use.  Computational performance using Python is achieved in UrbanSim and ActivitySim using vectorized calculations with math libraries such as Numpy that are implemented in C, thus avoiding the performance penalties of iterative processing in Python. 

\begin{figure}[htbp]
  \center
  \includegraphics[width=\textwidth]
  {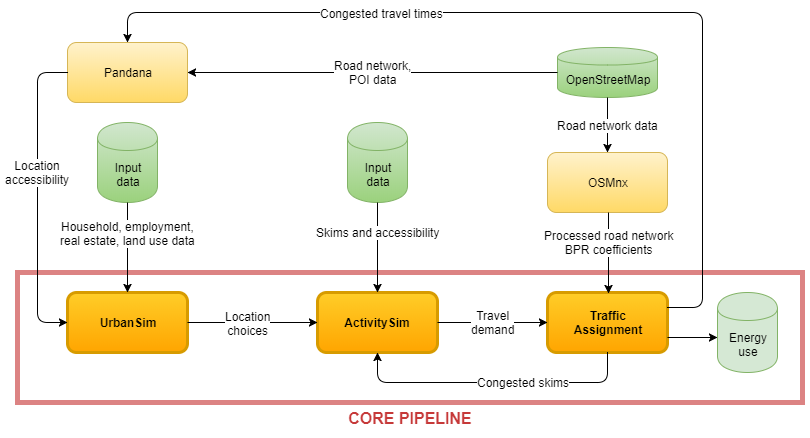}
  \caption{Overview of the integrated modeling pipeline's architecture}
  \label{fig:overview_pipeline_architecture}
\end{figure}

While UrbanSim and ActivitySim are microsimulation models, meaning that they operate at the level of individuals and households, the traffic assignment model is an aggregate traffic flow model that has been parallelized in a high-performance computing (HPC) environment. Figure \ref{fig:overview_pipeline_architecture} depicts the proposed pipeline for integrating these three models.

\section{Long-term land use model: UrbanSim}
\label{sec:urbansim}
\subsection{Overview}

UrbanSim has been developed to support land use, transportation and environmental planning, with particular attention to the regional transportation planning process \citep{waddell-japa-2002, waddell-tra-2007, waddell-tr-2011}. It has been designed to perform several tasks.

\begin{enumerate}
\item It can predict land use\footnote{We use the term \emph{land use} broadly, to represent the characteristics of real estate development and prices, and the location and types of households and businesses.} information for input to the travel model, for periods of 10 to 40 years into the future, as needed for regional transportation planning.

\item It can predict the effects on land use patterns from alternative investments in roads and transit infrastructure, alternative transit levels of service, or alternative roadway and transit pricing, over long-term forecasting horizons. Scenarios can be compared using different transportation network assumptions to evaluate the relative effects on development from a single project or a more wide-reaching change in the transportation system, such as extensive congestion pricing.

\item It can predict the effects of changes in land use regulations on land use. This includes the effects of policies to relax or increase regulatory constraints on development of different types, such as an increase in the allowed Floor Area Ratios (FAR) on specific sites, or allowing mixed-use development in an area previously zoned only for one use.

\item It can predict land use development patterns induced by investments in transit.

\item It can predict the effects of environmental policies that impose constraints on development, such as protection of wetlands, floodplains, riparian buffers, steep slopes, or seismically unstable areas.

\item It can predict the effects of changes in the macroeconomic structure or growth rates on land use. Periods of rapid or slow growth, or even decline in some sectors, can lead to changes in the spatial structure of the city and the model system is designed to analyze these shifts.

\item It can predict the possible effects of changes in demographic structure and composition of the city on land use and on the spatial patterns of clustering of residents of different social characteristics, such as age, household size, and income.

\item It can examine the potential impacts of major development projects (both actual and hypothetical) on land use and transportation. This can be used to explore the impacts of a corporate relocation or to compare alternative sites for a major development project.
\end{enumerate}

\subsection{Inputs}

UrbanSim requires detailed representation of the built environment and its occupants.  It uses a microsimulation data structure, meaning that it represents every household, person, job, parcel, and building in a metropolitan area. The population data is generated from census data using synthetic population algorithms we have developed \citep{ye-trb-2009}.  The employment data is from an inventory of business establishments from the Metropolitan Transportation Commission, as is the parcel and building data for the 9 county San Francisco Bay Area.

In addition, land use regulations from over 100 municipalities in the Bay Area were assembled and reconciled into a regional land use regulation database, and a database of development projects \enquote{in the pipeline} for development was also assembled, to be able to accurately reflect development that is in progress.  Real estate prices and rents were obtained from CoStar for use in estimating the rent and price models in UrbanSim.  Finally, the California Household Travel Survey was used in estimating the Household Location Choice Model in UrbanSim.

Beyond these base data sets, the other main input to UrbanSim is what we refer to as scenario inputs. We use the term \emph{scenario} in the context of UrbanSim in a very specific way: a scenario is a combination of input data and assumptions to the model system, including macroeconomic assumptions regarding the growth of population and employment in the study area, the configuration of the transportation system assumed to be in place in specific future years, and general plans of local jurisdictions that will regulate the types of development allowed at each location.

In order to facilitate comparative analysis, a model user such as a Metropolitan Planning Organization will generally adopt a specific scenario as a base of comparison for all other scenarios. This base scenario is generally referred to as the \emph{baseline} scenario, and this is usually based on the adopted or most likely to be adopted regional transportation plan, accompanied by the most likely assumptions regarding economic growth and land use policies. Once a scenario is created, it determines several inputs to UrbanSim:

\begin{itemize}
    \item \textit{Control totals}: data on the aggregate amount of population and employment, by type, to be assumed for the region.
    \item \textit{Travel data}: data on zone to zone travel characteristics, from the travel model.
    \item \textit{Land use plan}: data on general plans, assigned to individual parcels.
    \item \textit{Development constraints}: a set of rules that interpret the general plan codes, to indicate the allowed land use types and density ranges on each parcel.
\end{itemize}

\subsection{How it works}

\subsubsection{The Role of Accessibility}

Accessibility is a well explored area of urban theory, and is the concept that connects transportation and land use, so is is central to understanding the UrbanSim model system.  Kevin Lynch in \textit{Good City Form} states, \enquote{activities are assumed to locate according to the relative cost of reaching materials, customers, services, jobs, or labor.  Other values are simply subsidiary constraints in this struggle for access} \citep{lynch_good_1984}.  Lynch traces the original concepts to Wingo and Alonso \citep{wingo_transportation_1961,alonso_location_1964}, but the first explicit discussion is provided by Hansen \citep{hansen_how_1959}.

Operationally used accessibility frameworks include gravity-model based (defined by attractions and discounted by distance), cumulative-opportunity (summations within a set impedance measure) and space-time (limited by the opportunity prism of an individual's activity skeleton) \citep{kwan_space-time_1998,miller_measuring_1999}.  Dong and others expand on the space-time prisms by creating a logsum-based measure within a travel model \citep{dong_moving_2006}.  

To measure access, one must first choose a basic unit of geography to use.  The majority of transportation models in use today still rely heavily on zone-based geography for its simplicity and computational tractability \citep{hunt_current_2005}.  Zones can vary in size, but are usually a few city blocks at their smallest.  Drawbacks to this method include: the zones must be defined manually, they are arbitrary in scope, and they are too large to model micro-land use measures and walkability (which often vary on a block-by-block basis). We have developed algorithms to support walking scale queries on local street networks that enable fast accessibility queries to be computed on metropolitan scale networks \citep{foti2012generalized}.  In subsequent work we have generalized these algorithms to address walking plus transit networks \citep{blanchard-waddell-trr-2017, blanchard2017urbanaccess}.

\subsubsection{Model system design and geographic level of analysis}

UrbanSim has been adapted for use at differing levels of geography. In this project, we use a parcel-level specification of the data and models in UrbanSim to support the maximum level of geographic detail, to avoid imposing aggregation bias on the model system.

\begin{figure}[htbp]
    \center
    \includegraphics[width=\textwidth]{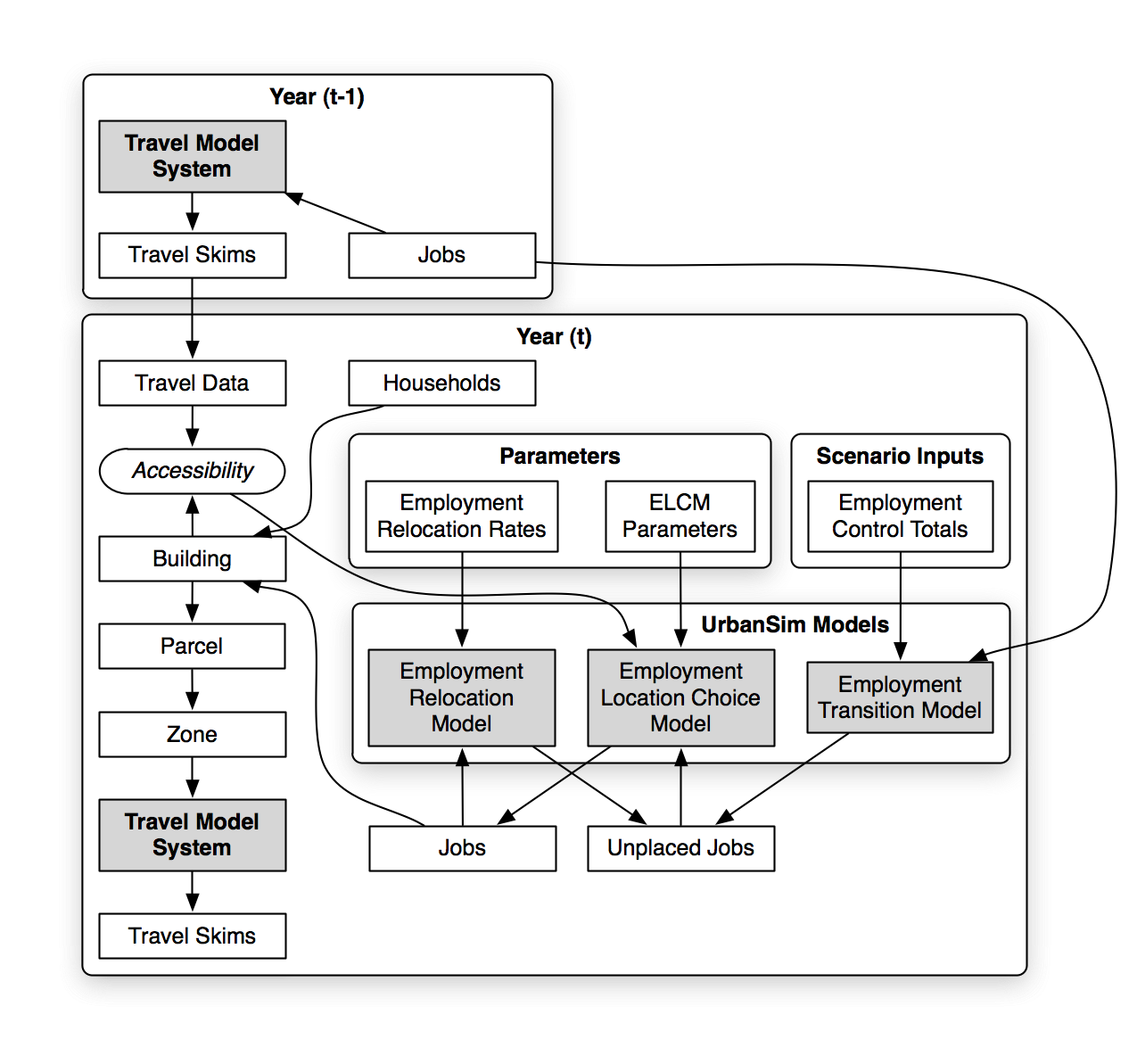}
    \caption{UrbanSim model flow: employment focus}
    \label{fig:employment-models}
\end{figure}

\begin{figure}[ht]
    \center
    \includegraphics[width=\textwidth]{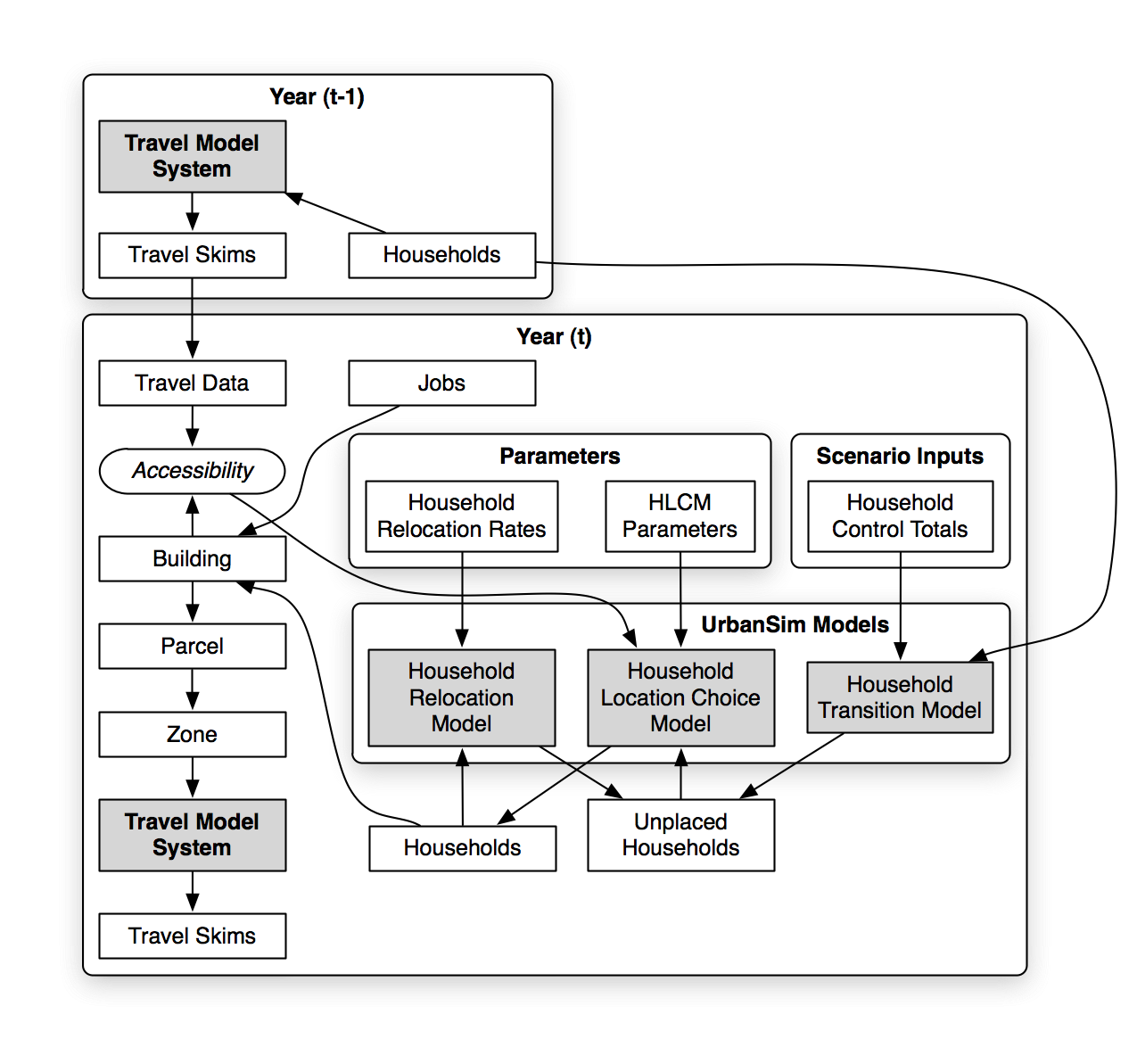}
    \caption{UrbanSim model flow: household focus}
    \label{fig:household-models}
\end{figure}

\begin{figure}[ht]
    \center
    \includegraphics[width=\textwidth]{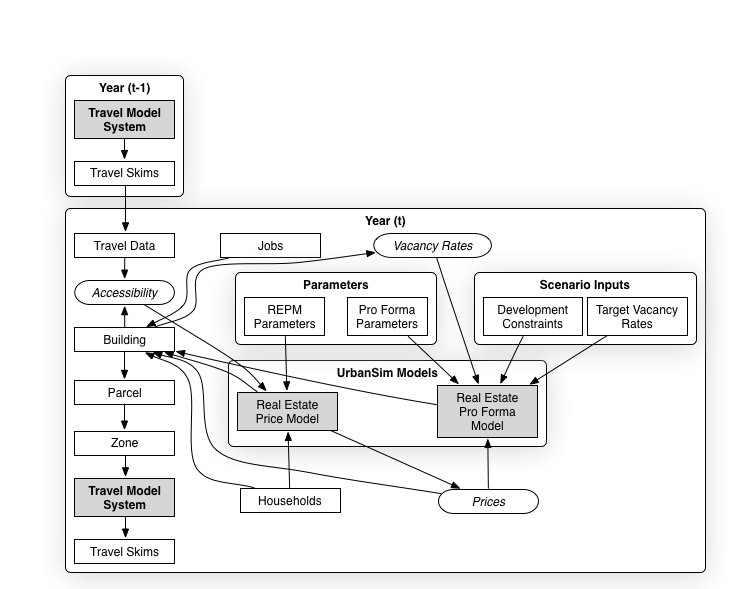}
    \caption{UrbanSim model flow: real estate focus}
    \label{fig:parcel-models}
\end{figure}

The components of UrbanSim are models acting on the objects in Figures \ref{fig:employment-models}, \ref{fig:household-models}, and \ref{fig:parcel-models}, simulating the real-world actions of agents in the urban system. Developers construct new buildings or redevelop existing ones. Buildings are located on land parcels that have particular characteristics such as value, land use, slope, and other environmental characteristics.

Governments set policies that regulate the use of land, through the imposition of land use plans, urban growth boundaries, environmental regulations, and through pricing policies such as development impact fees. Governments also build infrastructure, including transportation infrastructure, which interacts with the distribution of activities to generate patterns of accessibility at different locations that in turn influence the attractiveness of these sites for different consumers. Households have particular characteristics that may influence their preferences and demands for housing of different types at various locations. Businesses also have preferences that vary by industry and size of business (number of employees) for alternative building types and locations.

The model system contains a large number of components, so in order to make the illustrations clearer, there are three \enquote{views} of the system. In Figure \ref{fig:employment-models}, the focus is on the flow of information related to jobs. Figure \ref{fig:household-models} provides a household-centric view of the model system. Finally, Figure \ref{fig:parcel-models} provides a view with a focus on real estate.

UrbanSim predicts the evolution of these entities and their characteristics over time, using annual steps to predict the movement and location choices of businesses and households, the development activities of developers, and the impacts of governmental policies and infrastructure choices. The land use model is interfaced with a metropolitan travel model system to deal with the interactions of land use and transportation. Access to opportunities, such as employment or shopping, are measured by the travel impedance of accessing these opportunities via all available modes of travel.

\subsubsection{Discrete choice models}
\label{sec:discrete-choice}

UrbanSim makes extensive use of models of individual choice. A novel approach to modeling individual actions using discrete choice models emerged in the 1970s, with the pioneering work of McFadden on Random Utility Maximization theory \citep{mcfadden-1974,mcfadden-1981}. This approach derives a model of the probability of choosing among a set of available alternatives based on the characteristics of the chooser and the attributes of the alternative, and proportional to the relative utility that the alternatives generate for the chooser.

Maximum likelihood and simulated maximum likelihood methods have been developed to estimate the parameters of these choice models from data on revealed or stated preferences, using a wide range of structural specifications \citep{train-book-2003}. Early application of these models were principally in the transportation field, but also included work on residential location choices \citep{quigley-eer-1976,lerman-trr-1977,mcfadden-1978}, and on residential mobility \citep{clark-vanlierop-1986}.

Let us begin with an example of a simple model of households choosing among alternative locations in the housing market, which we index by $i$. For each agent, we assume that each alternative $i$ has associated with it a utility $U_i$ that can be separated into a systematic part and a random part:

\begin{equation}
    \label{eq:utility}
    U_i = V_i + \epsilon_i
\end{equation}

where $V_i = \beta\cdot {x}_i$ is a linear-in-parameters function, $\beta$ is a vector of $k$ estimable coefficients, $x_i$ is a vector of observed, exogenous, independent alternative-specific variables that may be interacted with the characteristics of the agent making the choice, and $\epsilon_i$ is an unobserved random term. Assuming the unobserved term in Equation \ref{eq:utility} to be distributed with a Gumbel distribution leads to the widely used multinomial logit model \citep{mcfadden-1974,mcfadden-1981}:

\begin{equation}
    \label{eq:mnl}
    P_i = \frac{\mathrm{e}^{V_i}}{\sum_j \mathrm{e}^{V_j}}
\end{equation}

where $j$ is an index over all possible alternatives. The estimable coefficients of Equation \ref{eq:mnl}, $\beta$, are estimated with the method of maximum likelihood \citep{greene-2002}.

The denominator of the equation for the choice model has a particular significance as an evaluation measure. The log of this denominator is called the \emph{logsum}, or composite utility, and it summarizes the utility across all the alternatives. In the context of travel mode between origins and destinations, for example, it would summarize the utility (disutility) of travel, considering all the modes connecting the origins and destinations. It has theoretical appeal as an evaluation measure for this reason. In fact, the logsum from the mode choice model can be used as a measure of accessibility.

\begin{figure}[htbp]
    \center
    \includegraphics[width=\textwidth]
    {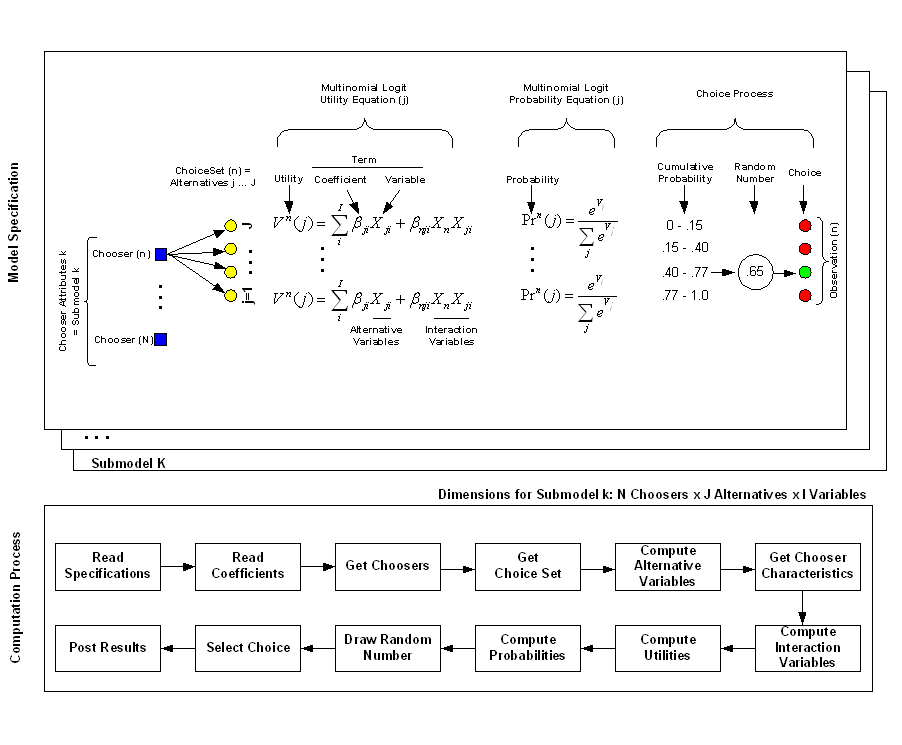}
    \caption{Computation process in UrbanSim choice models}
    \label{fig:choiceprocess}
\end{figure}

Choice models are implemented in UrbanSim in a modular way, to allow flexible specification of models to reflect a wide variety of choice situations. Figure \ref{fig:choiceprocess} shows the process both in the form of the equations to be computed and from the perspective of the tasks implemented as methods in software.

For each model component within the UrbanSim model system, the choice process proceeds as shown in Figure \ref{fig:choiceprocess}. The first steps of the model read the relevant model specifications and data. Then a choice set is constructed for each chooser. Currently this is done using random sampling of alternatives, which has been shown to generate consistent, though not efficient, estimates of model parameters \citep{ben-akiva-lerman-1987}.

The choice step in this algorithm warrants further explanation. Choice models predict choice probabilities, not choices. In order to predict choices given the predicted probabilities, we require an algorithm to select a specific choice outcome. A tempting approach would be to select the alternative with the maximum probability, but unfortunately this strategy would have the effect of selecting only the dominant outcome, and less frequent alternatives would be completely eliminated. In a mode choice model, for illustration, the transit mode would disappear, since the probability of choosing an auto mode is almost always higher than that of choosing transit. Clearly this is not a realistic outcome.

In order to address this problem, the choice algorithm used for choice models uses a sampling approach. As illustrated in Figure \ref{fig:choiceprocess}, a choice outcome can be selected by sampling a random number from the uniform distribution in the range 0 to 1, and comparing this random draw to the cumulative probabilities of the alternatives. Whichever alternative the sampled random number falls within is the alternative that is selected as the \enquote{chosen} one. This algorithm has the property that it preserves in the distribution of choice outcomes a close approximation of the original probability distribution, especially as the sample size of choosers becomes larger.

\begin{figure}[htbp]
    \center
    \includegraphics[width=\textwidth]
    {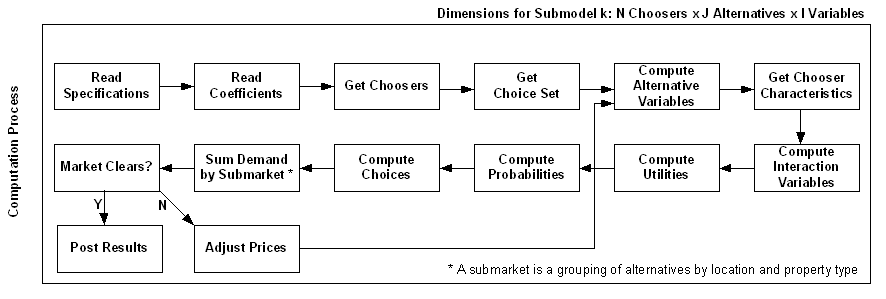}
    \caption{Choice process with price adjustment in UrbanSim choice models}
    \label{fig:choiceprocesswithprice}
\end{figure}

We have also developed an alternative choice algorithm that enables the model to simulate short-term market clearing processes. We compute the probability step of the location choice model, sum the probabilities at each submarket to compute aggregate demand, and use this estimate of demand to compare to the available supply in the submarket. Prices are adjusted iteratively and the relevant components of the location choice model are updated to reflect the influence of the adjusted prices. This algorithm captures the feedback loop between excess demand for locations causing prices there to increase, which in turn, dampens demand as the submarket becomes relatively more expensive than other submarkets that are substitutes. This variant of the choice process is shown in Figure \ref{fig:choiceprocesswithprice}.

One other choice context is worth noting. In some situations, the availability of alternatives may be constrained. If the limit on availability is entirely predictable, such as a budget constraint eliminating expensive options, or a zero-car household being unable to use the drive-alone mode, this is straightforward to handle, by eliminating the alternatives from the choice set for those choosers.

\subsection{Outputs}

As a microsimulation system, UrbanSim essentially produces the same outputs as it uses as inputs: tables of individual households and persons, jobs, parcels, buildings, with their attributes updated each simulation year if they have been modified by the model system.  New households, jobs and buildings are added or subtracted by the simulation, and households and jobs may relocate into or within the region.

By retaining this level of detail, UrbanSim is able to generate summaries of the real estate data, demographics, or economic profile of any geographic aggregation requested by the user, such as census geographies, cities, counties, or other planning geographies.  Often traffic analysis zone summaries are used as inputs to the travel model system.  In this project our goal is to avoid losing information by aggregating the data to traffic zone when we connect UrbanSim to the travel demand model system.

\subsection{Calibration and validation}

UrbanSim is generally calibrated longitudinally, starting from an observed year in the past and running it to a later observed year, to compare predicted to observed data over time, and adjusting calibration coefficients iteratively to improve the fit of the model to the observed calibration targets at the calibration year.  Generally it is preferable to minimize the use of calibration constants since excessive use of calibration constants can \enquote{handcuff} the model and make it insensitive to policy changes.

Some work has been done previously to extend this methodology to account for uncertainty using Bayesian Melding, to calibrate the model uncertainty and enable the computation of confidence intervals around its predictions, when running the model multiple times without fixing the random seed for the stochastic simulation \citep{sevcikova-tra-2009, sevcikova-tra-2011}.

\section{Short-term travel demand model: ActivitySim}
\label{sec:activitysim}
\subsection{Overview}

ActivitySim is an agent-based modeling (ABM) platform for modeling travel demand. Like UrbanSim, the ActivitySim software is entirely open source, and hosted as a part of the Urban Data Science Toolkit\footnote{The open-source Urban Data Science Toolkit is available online at \url{https://github.com/UDST}}. ActivitySim grew in large part out of a need for metropolitan planning organizations (MPOs) to standardize the modeling tools and methods that were common between them in order to facilitate more effective collaboration and sharing of innovations.

Today, ActivitySim is both used and maintained by an active consortium of MPOs, transportation engineers, and other industry practitioners. Because of the cooperative approach taken by ActivitySim stakeholders towards its ownership, and because many of its \enquote{owners} are also its main users, the platform continues to mature in the direction that most benefits the practitioners themselves. ActivitySim development is still in beta, with an official 1.0 release scheduled for 2018.

\subsection{Inputs}

ActivitySim requires two main sets of input data, one relating to geography and the other relating to the population of synthetic agents whose travel choices are being modeled.

The geographic data are stored at the level of the traffic analysis zone (TAZ) and are comprised of three components: 1) land use characteristics; 2) a matrix of zone-to-zone travel impedances (travel times, distances, or costs) specific to the mode of travel and time of day; and 3) a table of user-defined measures of aggregate utility estimated for each zone. In transportation planning, these zone-level impedances and utility measures are commonly referred to as \emph{skims} and \emph{accessibilities}, respectively.

The land use data consist of zone-level population and employment characteristics, along with measures of different land use and building types. In our integrated model these data are read directly from the outputs generated by UrbanSim, but for a single simulation iteration any source of aggregate land use data would suffice.

Travel skims are typically generated by a traffic assignment model, which ActivitySim is not. ActivitySim instead expects to load the skims from an OpenMatrix (OMX) formatted data file\footnote{The OpenMatrix format is specified online at \url{https://github.com/osPlanning/omx/wiki}}. The creation of these skims is described below in Section \ref{sec:ta} on traffic assignment.

Accessibilities can be generated directly from the skims or any other graph representation of the transportation network. They are computed by aggregating mode-specific measures of access to specific amenity types across the network, most commonly employment centers, retail outlets, and transportation hubs. The measures of access can be as simple as counts of amenities reachable within a given shortest-path distance or travel time, or as complex as composite utilities generated by a discrete choice model. 

The second set of ActivitySim input data is the synthetic population. The synthetic population data consist of both individuals and their characteristics, as well as the households and household characteristics into which the individuals are organized. The synthetic population is shared between UrbanSim and ActivitySim, although UrbanSim does not make use of individual-level characteristics.

The exhaustive details of the ActivitySim data schema are documented online\footnote{The ActivitySim data schema is available online at \url{https://udst.github.io/activitysim/dataschema.html}}.

\subsection{How it works}

ActivitySim, like UrbanSim, relies heavily on discrete choice models and random utility maximization theory \citep{mcfadden-1974}. Please refer to Section \ref{sec:urbansim} for specific details about how discrete choice models work within an agent-based microsimulation framework.

\begin{figure}[htbp]
    \center
    \includegraphics[width=\textwidth]{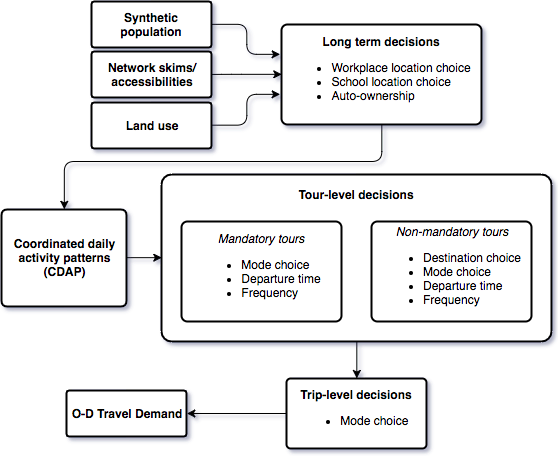}
    \caption[ActivitySim model flow]{ActivitySim model flow\footnote{The ActivitySim model flow is adapted from \url{http://analytics.mtc.ca.gov/foswiki/bin/view/Main/ModelSchematic}}}
    \label{fig:asim-models}
\end{figure}

An ActivitySim run consists of a series of sequentially executed model steps. The individual models can be grouped into the four clusters---long term decisions, coordinated daily activity patterns, tour-level decisions, and trip-level decisions---illustrated in Figure \ref{fig:asim-models} and summarized briefly here:

\begin{itemize}
    \item \emph{Long-term choice models}: ActivitySim's three long-term choice models---workplace location choice, school location choice, and auto-ownership---model the choices that are not made every day in the real world but have a substantial impact on those that are. These models will eventually be migrated to run directly in the UrbanSim environment so that the time horizons of the two simulation platforms are internally consistent.
    
    \item \emph{Coordinated Daily Activity Patterns}: the CDAP step models the group decision-making process for individual household members all seeking to maximize the utility of their daily activities together. CDAP takes into consideration mandatory and non-mandatory trips choosing activities to maximize each individual's utilities. The maximization process currently involves the estimation of all possible combinations of all individuals within a household, and thus has the longest run-time of all ActivitySim models.
    
    \item \emph{Tour-level decisions}: tours define chains of trips that are completed together without returning home in between. Mandatory tours include trips to and from work and school, while non-mandatory trips are entirely discretionary. Non-mandatory tour alternatives are specified in a user-defined configuration file, and thus these steps include a destination choice model as well. Mandatory tour alternatives have already been computed by the long-term decision models. Each tour type has separate model steps for estimating mode choice, departure time, and the frequency of the tour.
    
    \item \emph{Trip-level decisions}: mode choice must be selected at the level of the individual trip as well as the tour because a given tour may include different modes for different trip legs. Trip departure and arrival times are estimated as well. The rest of the trip characteristics are inherited from the tours to which a trip belongs. 
\end{itemize}

\subsection{Outputs}

The output of an ActivitySim run consists of a single HDF5 data file with a single table of results corresponding to each model step, along with the versions of the input files in their final, updated states. For the purpose of generating travel demand for traffic assignment, however, we are only concerned with the output of the trip generation step. This single file contains the origin and destination zones, start and end times, and mode choice for every trip taken by every agent over the course of a day. We then take the subset of these trips that are completed by automobile and aggregate the counts by origin-destination pair and hour of departure. These hourly, zone-level demand files are finally handed off for use in traffic assignment.

\subsection{Calibration and validation}
Compared to their meso- and macro-scale counterparts, microsimulations like ActivitySim more accurately capture the nonlinearities that define most patterns of human behavior by modeling the decision-making processes of individual agents. The models themselves, however, are not meant to be interpreted on the same disaggregate scale. We do not know which individuals will use which mode to complete which activity on a given day, but rather how an entire population of individuals is likely to behave \textit{en masse}.

As such, there are a variety of data sets available to us for validating our results, including the Bay Area Travel Survey (BATS), the U.S. Census Longitudinal Employer-Household Dynamics program (LEHD), and the California Household Travel Survey (CHTS). All of these products offer data that can be aggregated to the TAZ level and compared to the output of our models.

\section{Road network}
\label{sec:network}
\subsection{Overview}

Once we have produced this synthetic travel demand data, we model the regional circulation network for traffic assignment. Our integrated pipeline models networks as mathematical graphs consisting of a set of nodes \textit{N} connected to one another by a set of edges \textit{E} \citep{newman_networks:_2010,gastner_spatial_2006}. Specifically, the road network is modeled as a nonplanar directed multigraph with possible self-loops. The data come from OpenStreetMap.

This section describes how we acquire these data, construct a graph model of the network, process its topology, and calculate and impute relevant variables. Next it describes the process of calculating BPR coefficients and the assumptions baked into this calculation. Finally, it explains the process of converting the zone-based travel demand output from ActivitySim to a network node-based demand data set. 

\subsection{Network creation, construction, and processing}

The network data in this modeling pipeline come from OpenStreetMap. OpenStreetMap is a public worldwide collaborative mapping project and web platform with over two million users. Anyone may edit or access OpenStreetMap data, but community oversight and standards exist to prevent significant vandalism or inaccurate edits \citep{jokar_arsanjani_openstreetmap_2015}. In general, OpenStreetMap data are of high accuracy and quality, particularly in the United States and Western Europe \citep{corcoran_analysing_2013,over_generating_2010,haklay_how_2010,maron_how_2015}. OpenStreetMap imported the 2005 TIGER/Line roads data set as a foundation, and numerous corrections and additions to these data have been made since \citep{willis_openstreetmap_2008}.

\begin{figure}[htbp]
    \center
    \includegraphics[width=\textwidth]
    {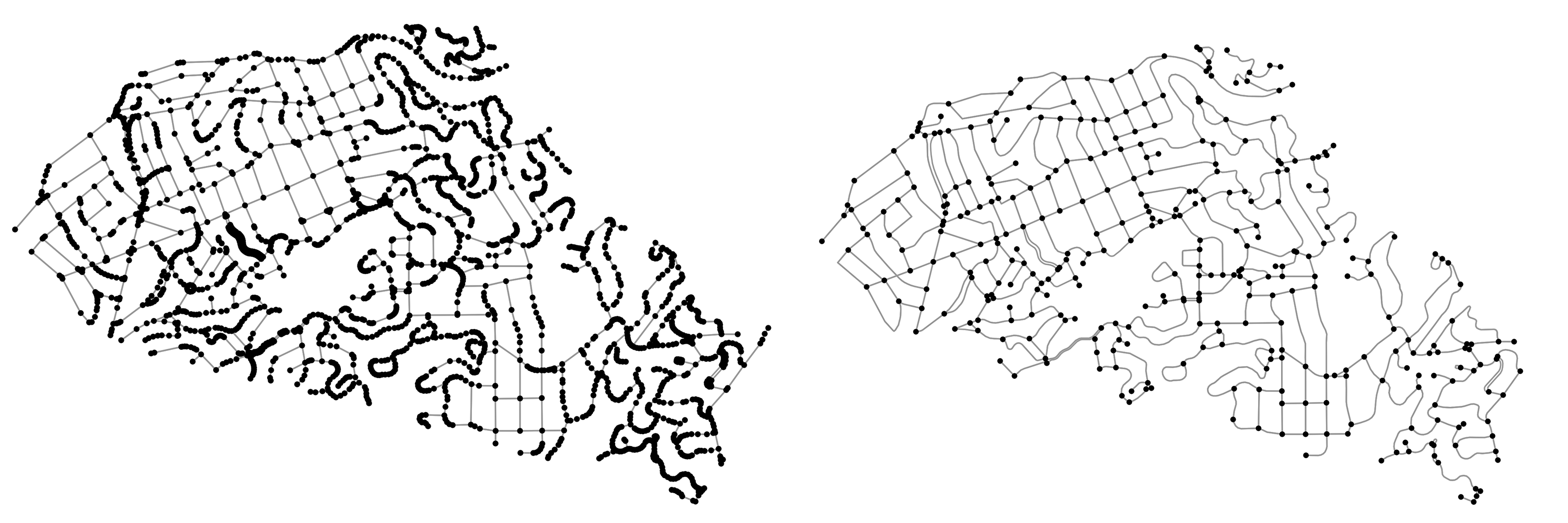}
    \caption{A suburban street network from OpenStreetMap data before OSMnx topology simplification (left) and after (right), with nodes in black and edges in gray. Note that the full spatial geometry is retained as edge metadata even though the edge itself is compressed to a single pair of origin and destination nodes.}
    \label{fig:simplification_before_after}
\end{figure}

The network graph is constructed using OSMnx, an open-source Python package for working with OpenStreetMap data \citep{boeing_osmnx:_2017}. OSMnx is built on top of NetworkX, a Python package for network analysis developed by researchers at Los Alamos National Laboratory. OSMnx extends NetworkX's network analysis capabilities by working explicitly with spatial infrastructure networks and interfacing with OpenStreetMap's various APIs \citep{boeing_methods_2017}. It can automatically construct topologically-corrected nonplanar directed multigraphs constrained to any polygonal boundaries for anywhere in the world from OpenStreetMap data. OSMnx uses a multistep algorithm to simplify the topology of the graph so that it retains nodes only at intersections and dead-ends, as well as the full geometry of the simplified edges, as shown in Figure \ref{fig:simplification_before_after}.

We use OSMnx to download the road network for the nine-county San Francisco Bay Area. We use the 2016 US Census Bureau TIGER/Line shapefile of United States counties to define the spatial extents of these nine counties: Alameda, Contra Costa, Marin, Napa, San Francisco, San Mateo, Santa Clara, Solano, and Sonoma. We calculate a convex hull around these geometries to obtain a single polygon for the spatial query---this prevents the query from discarding any network elements that fall within our study area but outside of a county's borders (for example, bridges over the San Francisco Bay).

Next, we use OSMnx to download the drivable road network within this convex hull. OSMnx processes the detailed metadata tags to identify which paths are drivable. It then constructs them into a graph. This initial graph contains 1.2 million nodes and 2.3 million edges.

Next we filter the road network to retain only tertiary roads and larger. In OpenStreetMap terminology, the road types we retain comprise: motorway, motorway\textunderscore link, trunk, trunk\textunderscore link, primary, primary\textunderscore link, secondary, secondary\textunderscore link, tertiary, tertiary\textunderscore link, unclassified, and road. The \enquote{link} types are necessary to retain the connectors (such as on-ramps and off-ramps) between certain roads. The \enquote{road} type is standardly used in the OpenStreetMap community as a null value. The \enquote{unclassified} type technically refers to the British-style roads hierarchy, in which \enquote{unclassified} is the level below \enquote{tertiary} – that is, quaternary. However, it sometimes is used inconsistently in the United States, so we retain it for completeness.

\begin{figure}[htbp]
    \center
    \includegraphics[width=\textwidth]
    {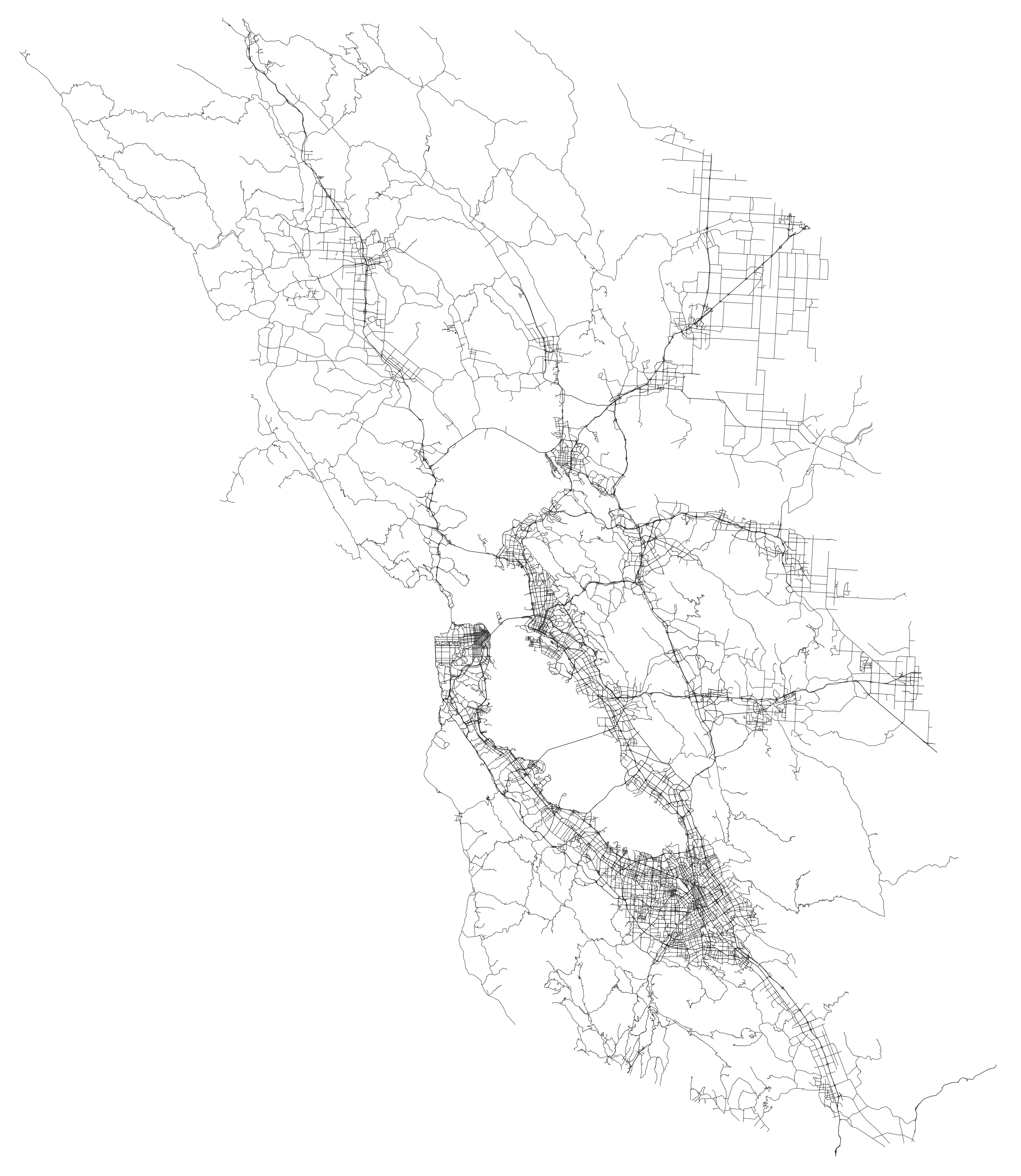}
    \caption{The nine-county San Francisco Bay Area road network used in our modeling pipeline, created by OSMnx from OpenStreetMap data.}
    \label{fig:bay_area_road_network}
\end{figure}

After filtering out all edges not of these major types, we remove all isolated nodes then keep only the largest weakly-connected component of the graph. Finally, we simplify the graph using OSMnx then save it to disk as a GraphML file. GraphML is a standard, XML-based file format for storing and exchanging complete graph structure data. Our final graph contains 31,000 nodes and 66,000 directed edges representing 20,000 kilometers of roads across these 9 Bay Area counties, as seen in Figure \ref{fig:bay_area_road_network}.

\subsection{BPR coefficients calculation and assumptions}
To define the relationship between edge travel time and edge congestion we use the Bureau of Public Roads (BPR) congestion function. This BPR function has long been used by transportation engineers to model the increased time needed to traverse an edge when congestion is high. Once we have the final processed graph of the Bay Area road network, we calculate the BPR  curves' coefficients $a_0$, $a_1$, $a_2$, $a_3$, and $a_4$ for it. The BPR function is defined in Equation \ref{eq:bpr_function}.

\begin{equation}
    t_i = t^0_i (1 + \alpha (\frac{v_i}{c_i}) ^ \beta)
    \label{eq:bpr_function}
\end{equation}

where:

\begin{itemize}
    \item $t_i$ is the congested flow travel time on edge $i$
    \item $t^0_i$ is the free-flow travel time on edge $i$
    \item $v_i$ is the number of vehicles on edge $i$ per unit of time 
    \item $c_i$ is the capacity (i.e., maximum number of vehicles) of edge $i$ per unit of time
    \item $\alpha$ linearly increases the congested travel time with regards to the volume:capacity ratio
    \item $\beta$ exponentially increases the congested travel time with regards to the volume:capacity ratio
\end{itemize}

The $\alpha$ parameter was assigned a value of 0.15 in the original BPR curve, and the $\beta$ parameter was assigned a value of 4 in the original BPR curve. We adopt these default $\alpha$ and $\beta$ parameter values in this integrated modeling pipeline, but note that they are not the only way to parameterize this function. For instance, the TRB's NHCRP Report 716: Travel Demand Forecasting Parameters and Techniques provides various coefficients estimated using the 1985 Highway Capacity Manual \citep[p.~75]{transportation_research_board_highway_1985,transportation_research_board_travel_2012}.

Given this function, we further define the coefficients $a_0$, $a_1$, $a_2$, and $a_3$ as:
 
\begin{itemize}
    \item $a_0 = t0_i$, that is, the free-flow travel time
    \item $a_1$, $a_2$, and $a_3$ are always zero
\end{itemize}

Then we calculate the coefficient $a_4$ as defined by Equation \ref{eq:bpr_a4_coefficient}:

\begin{equation}
    a_4 = t^0_i \frac{\alpha}{c_i ^ {\beta}}
    \label{eq:bpr_a4_coefficient}
\end{equation}

To calculate these coefficients, we require per-edge data about capacity and free-flow travel time. To assemble these data, we require information about edge lengths, free-flow speed, and number of lanes. We use OSMnx to calculate these edge lengths. OpenStreetMap contains sparse data per-edge on maximum permitted speed and number of lanes. When these data are missing, we infer or impute them as per the defaults\footnote{The authors wish to thank Madeleine Sheehan and Alexander Skabardonis for providing some of these values.} in Tables \ref{table:free_flow_speed_defaults} and \ref{table:capacity_defaults}.

First, for each edge that is missing number of lanes data, we impute the value based on its edge type (which OpenStreetMap always provides as metadata). We calculate this imputed value by taking the median value of all edges of this type. Second, for each edge that is missing maximum permitted speed data, we infer the free-flow speed from a look-up table via edge type and number of lanes (see Table \ref{table:free_flow_speed_defaults}). We can now calculate free-flow travel time per edge as a function of free-flow speed and edge length, as defined in Equation \ref{eq:free_flow_travel_time}:

\begin{equation}
    t^0_i = \frac{d_i}{s_i}
    \label{eq:free_flow_travel_time}
\end{equation}

where:

\begin{itemize}
    \item $t^0_i$ represents free-flow travel time on edge $i$, in units of seconds
    \item $d_i$ represents the length of edge $i$, in units of meters
    \item $s_i$ represents free-flow speed (i.e. the maximum permitted speed of travel) on edge $i$, in units of meters per second
\end{itemize}

Next, we infer each edge's vehicle capacity (per lane per hour) from a look-up table via edge type and number of lanes (see Table \ref{table:capacity_defaults}). We then convert this capacity per lane per hour value to units of capacity per edge per second. Finally, we use these data to calculate the values of the $a0$ and $a4$ coefficients for the BPR curve. These calculations demand a caveat: due to missing data on OpenStreetMap, we are forced to infer or impute variables on many edges. Our assumptions on parameter values and capacity and speed limit defaults propagate through to our BPR coefficients.

\begin{table}[htbp]
\centering
\caption{Free-flow speed defaults by edge type and number of lanes, in units of miles per hour.}
\label{table:free_flow_speed_defaults}
\begin{tabular}{ l r r r r } 
    \hline
    Edge type & lanes=1 & lanes=2 & lanes=3 & lanes=4+ \\
    \hline
    motorway & 50 & 50 & 65 & 65 \\
    motorway\textunderscore link & 50 & 50 & 65 & 65 \\
    trunk & 45 & 45 & 45 & 45 \\
    trunk\textunderscore link & 45 & 45 & 45 & 45 \\
    primary & 30 & 30 & 30 & 30 \\
    primary\textunderscore link & 30 & 30 & 30 & 30 \\
    secondary & 25 & 25 & 25 & 25 \\
    secondary\textunderscore link & 25 & 25 & 25 & 25 \\
    tertiary & 20 & 20 & 20 & 20 \\
    tertiary\textunderscore link & 20 & 20 & 20 & 20 \\
    unclassified & 20 & 20 & 20 & 20 \\
    road & 30 & 30 & 30 & 30 \\
    \hline
\end{tabular}
\end{table}

\begin{table}[htbp]
\centering
\caption{Capacity defaults by edge type and number of lanes, in units of vehicles per lane per hour.}
\label{table:capacity_defaults}
\begin{tabular}{ l r r r r } 
    \hline
    Edge type & lanes=1 & lanes=2 & lanes=3 & lanes=4+ \\
    \hline
    motorway & 1900 & 2000 & 2000 & 2200 \\
    motorway\textunderscore link & 1900 & 2000 & 2000 & 2200 \\
    trunk & 1900 & 2000 & 2000 & 2000 \\
    trunk\textunderscore link & 1900 & 2000 & 2000 & 2000 \\
    primary & 1000 & 1000 & 1000 & 1000 \\
    primary\textunderscore link & 1000 & 1000 & 1000 & 1000 \\
    secondary & 900 & 900 & 900 & 900 \\
    secondary\textunderscore link & 900 & 900 & 900 & 900 \\
    tertiary & 900 & 900 & 900 & 900 \\
    tertiary\textunderscore link & 900 & 900 & 900 & 900 \\
    unclassified & 800 & 800 & 800 & 800 \\
    road & 900 & 900 & 900 & 900 \\
    \hline
\end{tabular}
\end{table}

\subsection{Linking zone-based travel demand to the network}

The ActivitySim component of our integrated modeling pipeline produces an output dataset of zone-to-zone travel demand data. To model this travel demand (and in turn traffic assignment) on our road network, we must convert the trip origins and destinations from zones to network nodes.

We use the shapefile of Traffic Analysis Zones (TAZs) from the Bay Area Metropolitan Transportation Commission (MTC) to acquire zone spatial extents. Then we calculate the centroid of each zone polygon. Finally, we identify the network node nearest to each zone's centroid by taking the minimum of a vectorized calculation of great-circle distances from the centroid to every node in the network, using the haversine formula defined in Equation \ref{eq:haversine_formula}:

\begin{equation}
    \delta_{gc} = r\cdot \arccos{(\sin{\Phi_1}\cdot\sin{\Phi_2} + \cos{\Phi_1}}\cdot\cos{\Phi_2}\cdot\cos{|\lambda_1 - \lambda_2|})
    \label{eq:haversine_formula}
\end{equation}

where:

\begin{itemize}
    \item $\delta_{gc}$ represents the great-circle distance between the two points, in meters
    \item $r$ represents the radius of the Earth, in meters
    \item $\Phi_1$ and $\Phi_2$ represent the geographical latitudes of the two points, in radians
    \item $\lambda_1$ and $\lambda_2$ represent the geographical longitudes of the two points, in radians
\end{itemize}

Now that we have the road network, its per-edge BPR coefficients, and node-based travel demand data, we are ready to model traffic assignment.

\section{Traffic assignment model}
\label{sec:ta}
\subsection{Overview}

The traffic assignment component of this integrated modeling pipeline provides vehicles in the network specific paths to make trips between their origins and destinations (ODs). The origins, destinations, and numbers of vehicles making trips are received from ActivitySim and then assigned to specific network edges through traffic assignment. The resulting path assignment results in a total number of vehicles traveling on each edge in the transportation network, which is then used to compute a travel time on each edge. The travel time on an edge is a function of the number of vehicles using that edge. These travel times are then given back to UrbanSim and ActivitySim to be used for congested accessibility and skims.

\subsection{Inputs}

The primary inputs used by the traffic assignment model comprise the network (including the associated origin-destination travel demand) and BPR coefficients associated with each edge. These inputs were described in detail in Sections \ref{sec:activitysim} and \ref{sec:network} and reviewed briefly here.

\subsubsection{The network}

The network infrastructure is based on a set of nodes and edges in which connected edges share a node. Vehicles are allowed to travel from one edge to another if the edges are connected via a node. The ordered set of these edges (in which the ordering denotes a shared node between two edges) are called a path. In order for a vehicle to move from its origin to its destination, it must take a particular path. The possible paths between origins and destinations are defined by the topology of the network (i.e., the number of edges, the number of nodes, and the connections between these nodes and edges). The network described in Section \ref{sec:network} is used as the foundation of the traffic assignment model.

\subsubsection{BPR coefficients}

The BPR coefficients computed previously are used to assign a pseudo travel time to each edge as a function of the edge's load (i.e. number of vehicles on the edge). This is necessary in order to determine which paths vehicles should be assigned to (since vehicles are more likely to take a path which requires less time).

\subsection{How it works}

The currently implemented version of the traffic assignment model determines a static user equilibrium using the Frank-Wolfe algorithm. Static traffic assignment does not consider time varying parameters of any kind so there is no concept of flow dynamics but there is a well-defined equilibrium. Static user equilibrium, often called Wardrop's first principle in the transportation literature, is a well-defined state in which all vehicles take the shortest path from their origin to their destination. The resulting traffic assignment is based on shortest paths that are calculated while considering the travel times resulting from a loaded network in which each user is attempting to minimize their travel time. 

Wardrop's first principle states that the actual travel time experienced by a user in the network is equal or less than the travel time that the same driver would experience on any other route \citep{wardrop1952road}. Static user equilibrium---which is equivalent to Nash equilibrium---is defined in Equation \ref{eq:static_user_equilibrium}:

\begin{equation}
    \label{eq:static_user_equilibrium}
    \begin{split}
    \min \sum_{e\in\ \textit{edges}}\int_{0}^{v_e} S_e(x) dx\\
    \text{subject to} \\ 
    v_e = \sum_i \sum_j \sum_r \alpha_{ij}^{er} x_{ij}^{r}\\
    \sum_r x_{ij}^{r} = T_{ij}\\
    v_e \geq 0 \\
    x_{ij}^{r} \geq 0
    \end{split}
\end{equation}

The minimization is over travel time on each edge as a function of traffic volume. $x_{ij}^{r}$ is the number of vehicles on path $r$ from node $i$ to node $j$. $\alpha_{ij}^{ar}$ is equal to 1 if edge $a$ is contained in path $r$ and 0 otherwise. To perform this static user equilibrium calculation we use the Frank-Wolfe algorithm \citep{frank1956algorithm}. The Frank-Wolfe algorithm is a first-order optimization algorithm that is used to solve convex problems such as the static user equilibrium problem defined above. The algorithm works as follows; consider a problem of the form:

\begin{align*}
    S \in \mathcal{R}^n \text{ a polyhedron and } f: \mathcal{R}^n \rightarrow \mathcal{R} C^1, \text{ solve:}
\end{align*}

\begin{align*}
    \text{min. } f(x) \text{ subject to } x \in S
\end{align*}

Frank-Wolfe algorithm:

\begin{enumerate}
    \item Initialize with $x_0 \in S$ and let $k= 0$
    \item Determine a search direction $d_k = y_k - x_k$ by solving the linear program: \\
    $y_k \in \text{argmin}_{y \in S} \{ \nabla f(x_k)^Ty \} $
    \item Determine a step length $\alpha_k \in [0.1]$ such that: \\
    $f(x_k + \alpha_k d_k) = f((1 - \alpha)x_k + \alpha y_k) \leq f(x_k)$
    \item Update $x_{k+1} = (1 - \alpha)x_k + \alpha y_k$ let $k = k+1$ and go to step 2
\end{enumerate}

\subsection{Outputs}

From the traffic assignment model, we obtain the number of vehicles on each edge in the network and the resulting travel time (calculated using the BPR coefficients) of each edge. Since these travel times are a result of a static traffic assignment, they are not interpretable as exact travel times and the units are effectively meaningless. Rather, they represent the congested travel time of each edge calculated using the BPR equations.

As described previously, these BPR function are empirical. These travel times are used to generate congested skims and accessibility, in which the congestion on each edge in the network is known in relative terms, and therefore areas of relatively high versus low congestion can be specified. UrbanSim and ActivitySim use these congested edge travel times to calculate accessibility and skims. For these calculations, the precise definition of travel time is not necessary; the comparison between edges is the important component.

\subsection{Calibration and validation}

In its current implementation, the traffic assignment model is static, meaning that back-propagation of congestion and other time dependent phenomena observed in real transportation networks cannot be well-modeled. However, the resulting travel times due to loading on each edge can be compared with real data to understand how well the static approximation matches with the observed travel times on a network.

\section{Discussion and Conclusion}
\label{sec:conclusion}
\subsection{Summary of the project and architecture}

This technical report has presented the preliminary architecture of an integrated modeling pipeline that joins together long-term land use, short-term travel demand, and static user equilibrium traffic assignment.

Our land use model, UrbanSim, is an open-source microsimulation platform used by metropolitan planning organizations worldwide for modeling the growth and development of cities over long time horizons. Our travel demand model, ActivitySim, is an agent-based modeling platform that produces synthetic origin--destination travel demand data. Finally, we use a static user equilibrium traffic assignment model based on the Frank-Wolfe algorithm to assign vehicles to specific network paths to make trips between these origins and destinations. This traffic assignment model runs in a high-performance computing environment and the resulting congested travel time data can then be fed back into UrbanSim and ActivitySim for the next model run. 

This ongoing effort will focus on adding workplace choice and vehicle ownership into the UrbanSim models and tightly integrating this with network models. With the behaviorally integrated models and a sufficiently detailed representation of the transportation network and geography, ideally using local streets and parcels and buildings, we will be able to explore the long-term feedback effects of transportation infrastructure changes on urban development patterns. We will be able to include the feedback effects of these urban development dynamics on travel demand and on travel flows and speeds, and consequently on transportation-related energy consumption.

\subsection{Accomplishments to date}

In this preliminary phase of the project, we have recently completed an integration of our land use model (UrbanSim) and travel demand model (ActivitySim), which marks a first for urban microsimulation. We are constructing graph models of metropolitan-scale road networks on-demand, with configurable resolution (that is, tertiary roads and up, or all drivable roads if desired). Finally, we recently completed a preliminary hand-off of our synthetic travel demand data to a static user equilibrium traffic assignment model in an HPC environment, with a successful test model run.

This approach to integrated modeling offers several future benefits. It will allow for initial deliverables to be generated in the near term, e.g., a calendar year. The tools and analytics being used are robust and validated by previous research efforts and by operational applications among leading MPOs in the country. A community of users have already developed the necessary regional data for many urban environments, including the Bay Area, San Diego, Denver, Detroit, Seattle, etc. Finally, once developed, future efforts can begin to analyze more sophisticated policy scenarios.

By creating the combined models with sufficient performance to simulate the feedback of land use, travel, and congestion annually rather than every 5 to 10 years, a variety of alternative scenarios can be developed and explored that couple transportation and land use policies. Examples include transit-oriented development with transportation networks emphasizing transit, or more efficient highway projects coupled with supporting land use policies, with evaluation of the induced demand effects of both. Policies that affect parking construction and management, or which begin to explore the potential impacts of greater adoption of ride hailing services and eventually autonomous vehicles, could also be incorporated as data availability permits.

\subsection{Connections to energy use}

One objective of this work, in which we jointly consider long-term land use, short-term activity choices, and traffic assignment, is to observe the energy impacts that result as a consequence of particular decisions. Perhaps the most straightforward way to measure the energy impact of this system is to consider the emissions associated with transportation. Vehicular emissions can be calculated as a function of travel time \citep{ahn2008effects}. This is useful even with a static traffic assignment model in which the results are representative of travel time on each link and can therefore be used to calculate emissions costs \citep{aziz2012integration}.

As we move towards a dynamic traffic model, in which the results can be thought of as actual travel times and the effects of dynamic flow in the network are captured, these energy costs will become more precise.

\subsection{Open issues}

The main technical challenge currently preventing a deeper integration of our three modeling components relates to the issue of accessibilities and composite utilities. These measures are fundamental to the destination choice models of both UrbanSim and ActivitySim, and must be updated to reflect the latest network impedances as computed during the traffic assignment step. UrbanSim is configured to estimate its own accessibilities directly from a transportation network file using the Pandana Python package. ActivitySim, however, currently expects both an updated skims matrix and an associated set of destination logsums as inputs for its accessibility model step.

Because our travel model is not generating any composite utilities beyond travel times, we have no way of updating these ActivitySim inputs to incorporate the latest traffic assignment results for simulations beyond the first iteration. As such, a high priority for us is to reconfigure ActivitySim and all of its sub-models to run on the same Pandana-derived accessibility measures as UrbanSim. This will not only reduce the number of accessibility computations by a factor of two, but it will also significantly reduce the amount of overhead required for post-processing the traffic assignment output.

Another challenge related specifically to traffic assignment involves the transition from static to dynamic modeling. As described above, static user equilibrium is a well-defined mathematical notion in which there is a closed form solution and numerous algorithms exist to compute this solution. On the other hand, although many dynamic flow models exist, there is no well-defined notion of dynamic user equilibrium. Because of this, it is less clear whether the results of a dynamic traffic assignment are behaviorally realistic.

In order to validate the results from dynamic traffic assignment, these results must be compared with observed data. Although such a comparison can be used to gain an understanding of the performance of the dynamic model, there is no clear definition to understand exactly how well the dynamic assignment is performing.

\subsection{Future research agenda}

The upcoming research agenda for this project covers five primary areas: the network model, the travel demand model, traffic assignment, deeper integration of the pipeline, and validation/benchmarking.

\subsubsection{Network model}

We will be improving the network model in two primary ways. It currently includes tertiary roads and higher, discarding local and residential roads both for computational performance as well as because these types of edges historically have been unimportant to automobile-focused traffic assignment and congestion modeling. We will produce an alternative network model that includes all drivable edges to compare traffic assignment performance in the HPC cluster. Also, our zone-based synthetic travel demand data is currently mapped to the center-most node in each zone, from which all zone trips originate and to which all zone trips terminate. We will instead distribute zone-based origins and destinations according to a probability distribution across all the nodes in a zone.

\subsubsection{Travel demand model}

Our research agenda includes improvements to the travel demand model, ActivitySim. In addition to addressing the issue of accessibility calculations described in the previous section, we will be moving long-range models (including school choice, automobile ownership choice, and workplace choice) to UrbanSim, a more natural home for long-term decision modeling. Furthermore, we will research run-time improvements for ActivitySim to reduce its currently substantial execution time. We will also be working to more tightly couple the UrbanSim and ActivitySim runtime environments, eventually executing models from both platforms from within a single Python-based data pipeline orchestration framework. 

To increase the behavioral realism, we will simplify the behavior in the ActivitySim model system developed using the same software core as UrbanSim, and integrate the resulting models more directly into UrbanSim as a simplified, integrated, lightweight activity-based travel demand model system. This will then be tightly coupled to a distributed network flow model to simulate route choices, network loading, and generate updates of congested travel times. Using a parallel implementation of the network assignment model, our goal is to speed up the traffic assignment step dramatically, enabling it to be coupled to UrbanSim plus activity based travel demand model components that will run every simulated year.

In order to leverage application by MPOs, the approach we propose is to add functionality for including longer-term choices such as workplace, school location, and vehicle ownership to UrbanSim. This will significantly simplify the computation in ActivitySim and will enable a clear and viable path to achieving a long-term coupled urban dynamic and transport modeling framework that can effectively inform energy modeling via the rapid creation and evaluation of alternative scenarios. 

\subsubsection{Traffic assignment}

We have completed a preliminary traffic assignment on our network model with our travel demand data. Next we will model real-world scenarios using the San Francisco Bay Area land use data, travel demand data, and network model. Then we will be testing multiple scenarios, including networks with fewer or more edges, and signal timing. Finally we will be testing different traffic models, including Merchant Nemhauser, cellular automata, and link delay.

Initially, stochastic user equilibrium assignment models will provide the network flow modeling. Following this we propose to explore more robust network flow algorithms based on decomposition of convex programs. The network flow algorithms are easily decomposed for deployment in parallel computing environments (e.g., cloud and HPC). The parallelization of these algorithms will allow the travel demand and consequent network flow models to run fast enough to enable rapid integrated modeling with the UrbanSim framework. This can be handled by accelerated Frank-Wolfe algorithms that can run in a decentralized way on HPC platforms. More specifically, run times for UrbanSim for one simulation year in a large region such as the Bay Area is on the order of a few minutes, while simulating one day with the regional travel model requires many hours. Our goal is to reduce the combined model system to a run time under one hour per simulated year.

\subsubsection{Pipeline integration}

The current pipeline demonstrates a \enquote{loosest coupling} methodology wherein data inputs and outputs are serialized to disk for hand-off to the other modeling steps. Because UrbanSim and ActivitySim are both built on top of the orca model orchestration framework, we have been able to integrate their inputs and outputs. However, our research agenda includes automating the hand-off to the traffic assignment model, and pipelining the traffic assignment model's outputs as inputs that feed back into UrbanSim and ActivitySim to provide congested travel time information for location and travel decision making.

\subsubsection{Validation and benchmarking}

Finally, our research agenda includes substantial benchmarking and validation of this modeling pipeline. The benchmarking will measure run-time for various configurations of the models, with a research objective of improving run-time performance particularly for the travel demand model. We will also be exploring trade-offs between model granularity and representation versus computational performance. The validation step will compare our model's outputs with real-world data to verify plausible results.

Several MPOs would be in a position to evaluate and potentially apply this new framework, including MTC in the Bay Area, SANDAG in San Diego, PSRC in Seattle, and potentially SEMCOG in Detroit and DRCOG in Denver, as well as smaller MPOs such as PPACOG in Pikes Peak, NFRMPO in Colorado Springs, and OahuMPO in Honolulu---all of whom are currently using UrbanSim in their operational long-term transportation planning processes. We could draw on this experience to validate our approach to creating this simplified lightweight travel demand generator that is more appropriate for the time scales being modeled in the UrbanSim framework. 

\subsection{Deliverables}
This project has four key deliverables due to the Department of Energy over FY18. Our Q1 deliverable is this technical report on the progress so far and the preliminary architecture of this integrated modeling pipeline. The Q2 deliverable is a network flow model running at scale, integrated with UrbanSim and ActivitySim. The Q3 deliverable is a journal article manuscript describing these accomplishments, to be submitted for peer review. Finally, the Q4 deliverable is a code repository for this code to run at scale.

\clearpage
\addcontentsline{toc}{section}{Notes}
\theendnotes

\clearpage
\addcontentsline{toc}{section}{References}
\renewcommand\bibname{References}
\bibliographystyle{apalike}
\bibliography{references,refs-accessibility,refs-network}

\end{document}